# POLARIZATION OPTICS ANALOGY OF QUANTUM WAVEFUNCTIONS IN GRAPHENE

D. Dragoman – Univ. Bucharest, Physics Dept., P.O. Box MG-11, 077125 Bucharest, Romania, e-mail: danieladragoman@yahoo.com

**Abstract**

Detailed similarities between polarization states of light and ballistic charge carriers in graphene are derived. Based on these, the optical equivalent of quantum wavefunctions, Dirac equation, and the effect of an electrostatic potential are found, and the quantum analog of the refractive index of light and of the optical composition law of reflection coefficients are obtained. The differences between the behavior of quantum wavefunctions in graphene and electromagnetic fields, due to the chiral symmetry of ballistic charge carriers that cannot be mimicked in classical polarization optics, are also evidenced.

**Introduction**

There is a long tradition in establishing analogies between ballistic electrons and optical beams. These analogies are based on the formal similarities between the Helmholtz equation for electromagnetic waves and the time-independent Schrödinger equation describing the propagation of electron wavefunctions [1]. These similarities lead not only to the finding of sets of analogous parameters for different polarizations of the incident electromagnetic field, but also to the observation of similar phenomena, such as Bloch oscillations, the design of electron optics devices, such as lenses, prisms, and beam splitters, as well as the design of optical structures analogous to the quantum well, wires and dots encountered in the study of ballistic charge carriers (see the reviews in [2, 3]).

The discovery of graphene [4] brought the analogy between ballistic electrons and photons propagating in vacuum to a new level, since the dispersion relation of charge carriers in single-layer graphene is linear, just as that of photons, except that the light velocity in vacuum, $c$, is replaced by the Fermi velocity $v_F \cong c/300$ (see the recent reviews in [5, 6]). Moreover, it was shown that optical devices such as Fabry-Perot interferometers [7] and Bragg reflectors have counterparts in graphene if periodic electrostatic [8] or magnetic fields [9] are applied, and that in graphene specific phenomena to waveguide optics, such as the Goos-Hänchen effect, are present [10].

The aim of this paper is to further the already established analogies between charge carriers in graphene, which satisfy a Dirac-like equation and the polarization states of light. Quantum analogies to polarization states of light lead to the development of one of the first spintronic device: the Datta-Das transistor is equivalent to an electro-optic light modulator [11]. Such analogies could also prove fruitful in the case of graphene. More precisely, we show that propagation of charge carriers in graphene resembles that of polarization states in gyrotropic media that have electro-optical properties. As such, a deeper insight in charge carrier propagation in clean or disordered graphene could be obtained by optical experiments. We deal in this paper exclusively with single-layer graphene, which will be simply denoted as graphene in the following, and focus on the evolution of ballistic charge carriers.

The similarities emphasized in this paper between optical polarization states and charge carriers in graphene should not overshadow the different behavior of quantum wavefunctions and electromagnetic fields. For instance, for light waves propagating in homogeneous media, each component of the electromagnetic field satisfies the Helmholtz equation, and optical devices, such as polarizers and analyzers, can act independently on the *x* and *y* components of the electric field. In contrast, in graphene the two spinor components of the wavefunction cannot be changed independently: the charge carriers are chiral, i.e. they obey a charge-conjugation-like symmetry, which has no counterpart in classical optics.

**Optical analog of ballistic charge carriers**

The time-dependent Schrödinger equation could not be directly linked with the evolution of light waves since the time derivative in this equation is of first order, whereas a second-order time derivative appears in the Maxwell equation. However, the wavefunction of electrons that satisfy the time-independent scalar Schrödinger wave equation has been put into correspondence with one component of the electromagnetic field. In homogeneous graphene, on the other hand, the charge carriers are spinors that satisfy the time-independent Dirac equation [12]

$$\hbar v_F \boldsymbol{\sigma} \cdot \boldsymbol{k} \begin{pmatrix} \psi_1 \\ \psi_2 \end{pmatrix} = \hbar v_F \begin{pmatrix} 0 & k_x - ik_y \\ k_x + ik_y & 0 \end{pmatrix} \begin{pmatrix} \psi_1 \\ \psi_2 \end{pmatrix} = (E - V) \begin{pmatrix} \psi_1 \\ \psi_2 \end{pmatrix}, \qquad (1)$$

where $\boldsymbol{\sigma} = (\sigma_x, \sigma_y)$ consists of Pauli matrices, $\boldsymbol{k} = (k_x, k_y)$ is the wavevector of charge carriers, and (in homogeneous media) *V* is a constant potential energy. The dispersion relation of these charge carriers is then $E = V \pm |\hbar v_F k|$, the solution of (1) corresponding to positive (negative) energy being associated to electron (hole) states. Since the wavefunction of charge carriers is no longer scalar, its optical analog cannot be a certain component of the electromagnetic field, but an optical polarization wave. More precisely, in a homogeneous medium the solution of (1) is usually expressed as [12]

$$\psi = \begin{pmatrix} \psi_1 \\ \psi_2 \end{pmatrix} = \frac{1}{\sqrt{2}} \begin{pmatrix} 1 \\ s \exp(i\varphi) \end{pmatrix} \exp(ik_x x + ik_y y), \tag{2}$$

where $s = \text{sgn}(E - V)$, $\varphi = \tan^{-1}(k_y / k_x)$ and $k_x = \sqrt{(E-V)^2 / \hbar^2 v_F^2 - k_y^2}$. The quantum wavefunction in (2) is similar in form to the Jones vector

$$J = \begin{pmatrix} E_x \\ E_y \end{pmatrix} = \frac{1}{\sqrt{2}} \begin{pmatrix} 1 \\ \exp(i\theta) \end{pmatrix}, \tag{3}$$

which describes the polarization state of an electromagnetic wave with electric field components along the *x* and *y* axes $E_x$ and $E_y$, respectively. Depending on the value of $\theta$ the Jones vector in (3) represents an electromagnetic field linearly polarized at 45° or –45° (for $\theta = 0$ or $\pi$, respectively), a left or right circular polarized electromagnetic field (for $\theta = \pi/2$ or $3\pi/2$, respectively), or other generally elliptically polarized light waves.

The polarization state of light can be determined from interferometric and holographic measurements [13] or using space-variant subwavelength metal-stripe polarizers [14] in the spatial Fourier-transform polarimetry technique. Light polarization is essential in determining the dielectric constant of materials using ellipsometry [15], in polarization-sensitive optical coherence tomography [16], especially suited for examining biological samples, and in designing advanced devices such as polarization holographic gratings for fast optical recordings [17] and polarization-bistable lasers for optical memory [18].

In optics the *x* and *y* axes correspond to the optical axes of polarizer devices. To meaningfully define *x* and *y* axes for ballistic charge carriers that propagate in graphene, we should consider an interface between an ungated ($V = V_1 = 0$) and a gated ($V = V_2 = eV_b$, with $V_b$ the bias voltage applied on the gate) region, schematically represented in Fig. 1, the charge carriers being incident from the ungated region. In this case, the spinor wavefunction of normally incident electrons, for which $s = 1$, i.e. $\psi_+^T = 2^{-1/2}(1 \quad 1)$, (the superscript *T* indicates the transposition operation) has the same form as the Jones vector of an electromagnetic field linearly polarized at 45°, while the corresponding wavefunction for holes (with $s = -1$), $\psi_-^T = 2^{-1/2}(1 \quad -1)$, is similar to the Jones vector of an optical wave linearly polarized at −45°. In an analogous manner, the wavefunction for electrons (holes) that are tangent to the interface have the same form as the left (right) circular polarized electromagnetic field. For oblique incidence at an angle $\varphi$, the electron and hole wavefunctions, $\psi_+^T = 2^{-1/2}(1 \quad \exp(i\varphi))$ and $\psi_-^T = 2^{-1/2}(1 \quad -\exp(i\varphi))$, respectively, are orthogonal for any $\varphi$, since

$$(\psi_+^T)^* \cdot \psi_- = (\psi_-^T)^* \cdot \psi_+ = 0 \tag{4}$$

(* denotes complex conjugation). Orthonormal polarization states of light are important since they form a basis for any other polarization state. For ballistic charge carriers, however, such a decomposition of an arbitrary wavefunction in the basis states in (4) has no significance.

From (2) and (3) it follows that the time-independent propagation of charge carriers through a homogeneous medium can be seen as the electronic analog of an optical retarder applied on an electromagnetic field linearly polarized at 45°, the change of the optical wave through the action of the Jones matrix that characterizes the retarder, $M_\theta$, being

$$M_\theta \cdot \frac{1}{\sqrt{2}}\begin{pmatrix}1\\1\end{pmatrix} = \begin{pmatrix}1 & 0\\0 & \exp(i\theta)\end{pmatrix} \cdot \frac{1}{\sqrt{2}}\begin{pmatrix}1\\1\end{pmatrix} = \frac{1}{\sqrt{2}}\begin{pmatrix}1\\\exp(i\theta)\end{pmatrix}. \tag{5}$$

The phase difference introduced by the retarder between the *x*- and *y*-components of the electromagnetic field is similar to the incidence angle of ballistic charge carriers in graphene, and can be varied by changing this incidence angle. A variation of the angle of incidence leads to quantum wavefunctions that are similar to polarization states that vary from linearly polarized to circular polarized.

**Evolution law analogies**

It would be useful to extend the similarity between charge carrier states in graphene and polarization state in optics to the corresponding evolution equations, as for the optical analogy of ballistic electrons that satisfy the Schrödinger equation. In classical optics different components of the electric field that propagate in matter can be coupled by the tensor of dielectric constant in anisotropic materials, but this coupling does not involve generally first-order derivatives of the electric field, so that the Maxwell equation is different from the time-dependent Dirac equation. However, the spatial evolution law of the electric vector in anisotropic media, in the short wave approximation can be generally expressed as [19]

$$-i\frac{\lambda}{\pi}\frac{d}{dz}\begin{pmatrix}E_x\\E_y\end{pmatrix} = H_{opt}\begin{pmatrix}E_x\\E_y\end{pmatrix} = (\hat{\varepsilon}_r - n_0^2)\begin{pmatrix}E_x\\E_y\end{pmatrix} = \begin{pmatrix}\alpha & \beta+i\gamma\\\beta-i\gamma & -\alpha\end{pmatrix}\begin{pmatrix}E_x\\E_y\end{pmatrix}, \tag{6}$$

where $\hat{\varepsilon}_r$ is the relative dielectric tensor when external fields that induce the anisotropy are applied and $n_0$ is the refractive index of the isotropic medium. Equation (6) is valid for monochromatic electromagnetic waves that propagate with wavelength $\lambda$ in the medium along the $z$ axis, which is one of the principal axes of the dielectric tensor (in [19] equation (6) is written for the displacement vector, but it holds also for the electric field). Therefore, $H_{opt}$ describes a similar spatial evolution of the polarization state as the time evolution of the quantum wavefunction in graphene under the Hamiltonian $H = \hbar v_F \boldsymbol{\sigma} \cdot \boldsymbol{k} + V$ if $V = \alpha = 0$ and if the replacements $2\lambda \leftrightarrow \hbar$, $z \leftrightarrow t$, $E_x \leftrightarrow \psi_1$, $E_y \leftrightarrow \psi_2$, $\alpha \leftrightarrow k_x$, $\beta \leftrightarrow -k_y$, are performed.

The question is if there is a medium in which the degree of anisotropy $\Delta\hat{\varepsilon} = \hat{\varepsilon} - n_0^2$ can be expressed by the matrix

$$\Delta\hat{\varepsilon} = \begin{pmatrix} 0 & \beta + i\gamma \\ \beta - i\gamma & 0 \end{pmatrix}. \tag{7}$$

The $\gamma$ parameter is finite in gyrotropic media. These are either optically active (or chiral media), in which $\gamma$ is the wavevector-dependent gyration vector, or magneto-optic materials, in which $\gamma$ depends on a uniform magnetic field applied on the otherwise isotropic material in the Faraday configuration [20]. Chiral media obey the reciprocal principle in optics, whereas magneto-optical media do not [21]. Optically active materials are, among others, quartz, AgGaSe$_2$, Se and Te, while magneto-optic materials include water, diamond and glass [22]. In a gyrotropic medium, the $\beta$ parameter can be controllable induced, for example, by an electric field applied along $z$ and is related to the electro-optic coefficient in materials with cubic $\bar{4}3m$ or tetragonal $\bar{4}2m$ crystalline structures; such materials are, such as KDP, AgGaSe$_2$, GaAs, CdTe, GaP, ZnSe and ZnTe [22]. A comparison between the gyrotropic and electro-optic materials shows only one overlap: AgGaSe$_2$, but it is also possible to engineer materials that are both gyrotropic and electro-optic. One method is to fabricate superlattices composed of alternating layers of gyrotropic and electro-optic materials. In this way, the total dielectric constant tensor has contributions from the two types of media [23].

The identification of the similarity between polarization states of light and charge carriers in graphene allows us to address the problem of the significance of the quantum analog of the Stokes parameters [20]. After straightforward calculations we find that for the optical-graphene analogy, $S_0 \leftrightarrow |\psi|^2 = 1$ is the density of charge carriers, $S_1 = 0$, $S_2 \leftrightarrow j_x = s\cos\varphi$, and $S_3 \leftrightarrow j_y = s\sin\varphi$, where $j_{x,y} = (\psi^T)^* \cdot \sigma_{x,y} \psi$ are the components of the electrical currents on $x$ and $y$ axes. It can be seen that, as for polarized waves in optics, in the quantum case we have also $S_1^2 + S_2^2 + S_3^2 = S_0^2$.

**Optical analog of electrostatic gates on graphene**

To find the optical analog of a gated region, in which the potential $eV_b$ (see Fig. 1) can be modified by an applied voltage on the gate, we must first solve the Dirac equation (1). The wavefunction can be written as [12]

$$\psi_1(x,y) = \exp(ik_y y) \times \begin{cases} \exp(ik_1 x) + r_{12}\exp(-ik_1 x), & x < 0 \\ t_{12}\exp(ik_2 x), & x \geq 0 \end{cases} \quad (8a)$$

$$\psi_2(x,y) = \exp(ik_y y) \times \begin{cases} s_1[\exp(ik_1 x + i\varphi_1) - r_{12}\exp(-ik_1 x - i\varphi_1)], & x < 0 \\ s_2 t_{12}\exp(ik_2 x + i\varphi_2), & x \geq 0 \end{cases} \quad (8b)$$

where $s_1 = \text{sgn}(E - V_1)$, $\varphi_1 = \tan^{-1}(k_y / k_{x1})$, $k_{x1} = \sqrt{(E - V_1)^2 / \hbar^2 v_F^2 - k_y^2}$, $s_2 = \text{sgn}(E - V_2)$, $\varphi_2 = \tan^{-1}(k_y / k_{x2})$, $k_{x2} = \sqrt{(E - V_2)^2 / \hbar^2 v_F^2 - k_y^2}$, and $r_{12}$ and $t_{12}$ are the reflection and transmission coefficients for charge carriers propagating from region 1 to region 2.

Imposing the continuity condition at the interface $x = 0$ for the two spinor components in (8), we obtain the reflection and transmission coefficients:

$$r_{12} = \frac{s_1 \exp(i\varphi_1) - s_2 \exp(i\varphi_2)}{s_1 \exp(-i\varphi_1) + s_2 \exp(i\varphi_2)} = \frac{s_1(k_{x1} + ik_y)/k_{F1} - s_2(k_{x2} + ik_y)/k_{F2}}{s_1(k_{x1} - ik_y)/k_{F1} + s_2(k_{x2} + ik_y)/k_{F2}}, \quad (9a)$$

$$t_{12} = \frac{2s_1 \cos\varphi_1}{s_1 \exp(-i\varphi_1) + s_2 \exp(i\varphi_2)} = \frac{2s_1 k_{x1}/k_{F1}}{s_1(k_{x1} - ik_y)/k_{F1} + s_2(k_{x2} + ik_y)/k_{F2}}. \quad (9b)$$

where $k_{F1,2} = (E - V_{1,2})/\hbar v_F$ are the Fermi wavenumbers in the two regions. As in optics, $R + T = 1$, where $R = |r_{12}|^2$, $T = (s_2 \cos\varphi_2 / s_1 \cos\varphi_1)|t_{12}|^2$.

However, despite the optical-graphene similarities evidenced in the previous section, the form of the reflection and transmission coefficients in (9) differ from the corresponding expressions for light waves normally incident on an interface between regions with (possibly complex) refractive indices $n_1$ and $n_2$:

$$r_{12}^{em} = \frac{n_1 - n_2}{n_1 + n_2}, \quad t_{12}^{em} = \frac{2n_1}{n_1 + n_2}. \quad (10)$$

(We consider in this section the analogy with normally incident electromagnetic waves since in this case both $E_x$ and $E_y$ are tangent to the interface and thus satisfy continuity condition.) An example of this difference is that, at normal incidence $r_{12} = 0$ if $s_1 = s_2$ (the electron- or hole-like character of the charge carriers is preserved), irrespective of the potential energies in the two media, whereas for electromagnetic waves $r_{12}^{em} \neq 0$.

The difference between the expressions of reflection and transmission coefficients in the quantum and optical cases is not related to boundary conditions, since both $x$ and $y$ components of the electric field are continuous for a normally incident wave, as the spinor components for graphene. To incorporate the phase factors $\varphi_1$ and $\varphi_2$ in the expressions of $E_y$ in the optical case we must add to the abrupt interface infinitely thin birefringent layers that introduce opposite phase differences between electromagnetic waves components that propagate in opposite directions (see Fig. 2). More precisely, on each side of the interface the $E_y$ component should acquire an extra phase (relative to $E_x$) of $\theta_{1,2}$ if propagating in one direction and $-\theta_{1,2}$ if its direction changes, so that the conditions at the quantum interface are replaced by continuity conditions on different planes in the optical case: $x = 0$, $0_-$, and $0_+$, respectively, for the incident, reflected and transmitted waves. This is quite an awkward situation, which only emphasizes the different behavior of quantum wavefunctions and electromagnetic fields. In particular, for light waves propagating in homogeneous media, polarizers or analyzers can modify independently the $x$ and $y$ components of the electric field. In contrast, in graphene the two spinor components of the wavefunction cannot be changed independently since the chirality of charge carriers has no counterpart in classical optics.

From equations (8) it follows that the transport of charge carriers across the interface is also formerly equivalent to a change in the polarization state of an incident electromagnetic field. Disregarding the reflected wave and the amplitude modulation of the transmitted wave, such a polarization change between the transmitted spinor $\psi_t^T = 2^{-1/2}(1 \quad s_2 \exp(i\varphi_2))$ and the incident one, $\psi_{in}^T = 2^{-1/2}(1 \quad s_1 \exp(i\varphi_1))$, can be seen as induced by the quantum equivalent of an optical phase retarder. The matrix $M$ of this device, determined from $\psi_t = M \psi_{in}$,

$$M = \begin{pmatrix} 1 & 0 \\ 0 & \mathrm{sgn}(s_2/s_1)\exp[i(\varphi_2 - \varphi_1)] \end{pmatrix}, \qquad (11)$$

is indeed similar to the matrix of a phase retarder,

$$M_{ret} = \begin{pmatrix} 1 & 0 \\ 0 & \exp(i\Delta\theta) \end{pmatrix}, \tag{12}$$

which connects the transmitted and incident Jones polarization vectors according to $J_t = M_{ret} J_{in}$ and delays one polarization component with respect to the other. The action of the two devices are the same if $\Delta\theta \leftrightarrow (s_2/s_1)\exp[i(\varphi_2 - \varphi_1)]$. The quantum equivalent phase shift of an optical retarder can thus be modified by changing the bias applied on the gate and/or the angle of incidence of ballistic charge carriers in graphene. In optics, phase retarders for optical waves of wavelength $\lambda$ can be implemented with birefringent materials, the phase delay $\Delta\theta = (2\pi/\lambda)\Delta n d$ between the electric field components being proportional to the product of the material birefringence $\Delta n$ and the thickness of the retarder, $d$. So, the variable electrostatic field applied through the gate bias has a similar effect on the wavefunction in graphene as an optical retarder with either a variable birefringence or a variable thickness.

Because the interface between regions with different potential energies in graphene leads not only to a change of propagation direction, i.e. refraction, of the wavefunction of charge carriers, but also to a change in the polarization state of the corresponding Jones vector, its optical analog is a polarizing beam splitter.

**Snell law for ballistic charge carriers**

Refraction of optical waves at an interface that separates two media with refractive indices $n_1$ and $n_2$ is governed by the Snell law $n_1 \sin\theta_1 = n_2 \sin\theta_2$, where $\theta_i$, $i = 1,2$, are the angles between the propagation direction in the two media and the normal to the interface. A similar relation can be found also for the refraction of electron wavefunctions at the interface between regions with different potential energies in graphene. In this case, the equality of the tangential wavevector at the interface imposes that

$$k_{x1} \tan \varphi_1 = k_{x2} \tan \varphi_2 \qquad (13)$$

or, taking into account the dispersion relation $k_{xi} = \sqrt{(E-V_i)^2/\hbar^2 v_F^2 - k_y^2}$,

$$\left(\frac{E-V_1}{\hbar v_F}\right)\sin\varphi_1 = \left(\frac{E-V_2}{\hbar v_F}\right)\sin\varphi_2. \qquad (14)$$

Equation (14) represents the Snell law for ballistic charge carriers in graphene, which is similar to that for electromagnetic fields if the refractive index is replaced by $(E-V)/\hbar v_F$.

The Snell law in (14) shows that in passing from electron-like (*n*-type) states to hole-like states or vice-versa at an interface between regions with different potential energies in graphene, the sign of the refraction angle changes. This metamaterial-like behavior of *p-n* junctions has been predicted in [24], where the possibility of implementing a Veselago lens in graphene has been discussed. We would like to point out here that formula (14) shows that an equivalent negative refractive index implies a negative normal wavevector component, which is not consistent with the expression of the wavefunction of ballistic charge carriers in (8), but is consistent with the corresponding expression for hole states derived in [25]. In the latter case, the *x* component of the wavevector for hole states changes direction (see also [24]), and the corresponding Jones vectors for electron and hole states incident under the same angle,

$$\psi'_+ = \frac{1}{\sqrt{2}}\begin{pmatrix} 1 \\ \exp(i\varphi) \end{pmatrix}\exp(ik_x x), \quad \psi'_- = \frac{1}{\sqrt{2}}\begin{pmatrix} 1 \\ \exp(-i\varphi) \end{pmatrix}\exp(-ik_x x) \qquad (15)$$

(the common factor $\exp(ik_y y)$ is disregarded) are complex conjugate, i.e. $\psi'_- = \psi'^*_+$. For these spinor wavefunctions, the reflection and transmission coefficients at the interface between media 1 and 2 when the incident electron states transform in hole states become

$$r'_{12} = \frac{\exp(i\varphi_1) - \exp(-i\varphi_2)}{\exp(-i\varphi_1) + s_2 \exp(-i\varphi_2)}, \qquad t'_{12} = \frac{2\cos\varphi_1}{\exp(-i\varphi_1) + \exp(-i\varphi_2)}. \tag{16}$$

These expressions are finite even at normal incidence, whereas the reflection and transmission coefficients in (9) take singular values at normal incidence if $s_1 = -s_2$, i.e. if electron states transform into hole states on the other side of the interface. Thus, the optical-graphene analogy is proving useful also in establishing the appropriate form of the graphene wavefunction when both electron and hole states are involved.

**Composition law of reflection coefficients**

The composition law of optical reflection coefficients in multilayer media has been shown to be similar to the law of velocity composition in special relativity. A similar composition law is expected to hold for the reflection coefficient of the wavefunction of charge carriers in graphene for two reasons: (i) the dispersion law is similar to that of photons, and (ii) the charge carriers in graphene are ultrarelativistic.

Let us consider now that charge carriers in graphene propagate through the region labeled by 2 and are observed in region 3, in which the potential energy is $V_3$. In this case, if the two interfaces are located at $x = x_1$ and $x = x_2$, so that region 2 has a width $x_2 - x_1 = L$, the spinor components of the wavefunction in the two regions can be expressed as [12]

$$\psi_1(x,y) = \exp(ik_y y) \times \begin{cases} \exp(ik_1 x) + r\exp(-ik_1 x), & x < x_1 \\ A\exp(ik_2 x) + B\exp(-ik_2 x), & x_1 \leq x < x_2 \\ t\exp(ik_3 x), & x \geq x_2 \end{cases} \tag{17a}$$

$$\psi_2(x,y) = \exp(ik_y y) \times \begin{cases} s_1[\exp(ik_1 x + i\varphi_1) - r\exp(-ik_1 x - i\varphi_1)], & x < x_1 \\ s_2[A\exp(ik_2 x + i\varphi_2) - B\exp(-ik_2 x - i\varphi_2)], & x_1 \leq x < x_2 \\ s_3 t\exp(ik_3 x + i\varphi_3), & x \geq x_2 \end{cases} \tag{17b}$$

where $s_3 = \text{sgn}(E - V_3)$, $\varphi_3 = \tan^{-1}(k_y/k_{x3})$, $k_{x3} = \sqrt{(E-V_3)^2/\hbar^2 v_F^2 - k_y^2}$, $r$ and $t$ are the reflection and transmission coefficients of the entire structure, and the other parameters are defined as before. Imposing the appropriate boundary conditions at the interfaces we can express the reflection coefficient of the entire structure in terms of the reflection coefficients at the two boundaries, $r_{12}$ in (9a) and

$$r_{23} = \frac{s_2 \exp(i\varphi_2) - s_3 \exp(i\varphi_3)}{s_2 \exp(-i\varphi_2) + s_3 \exp(i\varphi_3)} \tag{18}$$

as

$$r_{13} = \frac{r_{12} + r_{23} \exp(2i\alpha_2) \exp(i\Phi)}{1 + r_{12}^* r_{23} \exp(2i\alpha_2) \exp(i\Phi)} \exp(2i\alpha_1), \tag{19}$$

where $\alpha_j = k_j(x_j - x_{j-1})$ with $x_0 = 0$, and

$$\Phi = \text{Arg}\left[\frac{s_1 \exp(i\varphi_1) + s_2 \exp(-i\varphi_2)}{s_1 \exp(-i\varphi_1) + s_2 \exp(i\varphi_2)}\right] \tag{20}$$

with Arg[] denoting the argument of a complex number of unit modulus (in our case).

The expression of the total reflection coefficient of ballistic electrons is different but similar to the corresponding formula for electromagnetic waves that traverse under an angle $\theta_2$ a slab of width $L$ and refractive index $n_2$, placed between regions with refractive indices $n_1$ and $n_3$:

$$r^{em} = \frac{r_{12}^{em} + r_{23}^{em} \exp(i2\beta_2)}{1 + r_{12}^{em} r_{23}^{em} \exp(2i\beta_2)} \exp(i2\beta_1), \tag{21}$$

$\beta_j = (2\pi/\lambda)n_j \cos\theta_j(z_j - z_{j-1})$. The two interfaces are again located at $x = x_1$ and $x = x_2$, with $x_2 - x_1 = L$, and $x_0 = 0$. For electromagnetic waves the reflection coefficients $r_{12}^{em}$ and $r_{23}^{em}$ depend on polarization and are given by (10) for normal incidence only. Both quantum and optical composition laws of reflection coefficients can be generalized to several layers.

Because the composition law in (21), written under the form

$$R = R_1 \oplus R_2 = \frac{R_1 + R_2}{1 + \overline{R}_1 R_2} \tag{22}$$

where $R = r/A$, with $A$ the incident amplitude, $R_j = (r_{j,j+1}^{em}/A)\exp[i2(\beta_1 + ... + \beta_j)]$, $\overline{R}_j = (r_{j,j+1}^{em}/A)\exp[-i2(\beta_1 + ... + \beta_j)]$, is similar to the Einstein addition law of parallel velocities $v_1$ and $v_2$ in special relativity,

$$v = v_1 \oplus v_2 = \frac{v_1 + v_2}{1 + v_1 v_2/c^2}, \tag{23}$$

stratified planar structures can be used to simulate the relativistic addition of parallel velocities [26]. Formula (22) is also used to calculate energy levels and wavefunctions in quantum potentials of arbitrary shapes [27]. A generalization of this optical-relativity similarity has been found for absorbing optical multilayer structures [28] and for non-parallel boosts [29]. These results are not unexpected since the Jones matrix formalism has been shown to be a representation of the six-parameter Lorentz group [30], the Stokes parameters forming a Minkowski four-vector [31], so that the Wigner rotation in special relativity can be mimicked with optical devices [32]. A review on the classical optical representations of the Lorentz group is found in [33].

The composition law of reflection coefficients in graphene, (19), is similar to that in multilayer optics, (22), if $R_1 = (r_{12}/A)\exp(i2\alpha_1)$, $\overline{R}_1 = (r_{12}^*/A)\exp(-i2\alpha_1)$, and $R_2 = [r_{23}\exp(i\Phi)/A]\exp[i2(\alpha_1+\alpha_2)]$. It follows then that graphene can also simulate the relativistic addition law of boosts, application that expands the graphene-relativity similarities to the case of special relativity; quantum electrodynamics phenomena such as the Klein paradox can also be studied in graphene [12, 25, 34].

**Quantum analog of the geometric phase factor in optical multilayer structures**

In an optical multilayer structure it was shown that a geometric phase factor appears due to the noncommutativity of the addition law for the reflection coefficient. More precisely, in the particular case of light propagation through a slab, studied in the previous section, it is easy to shown that [35]

$$R_1 \oplus R_2 = (R_2 \oplus R_1)\exp(i2\vartheta), \tag{24}$$

with

$$\vartheta = \mathrm{Arg}\left(\frac{1+R_1\overline{R}_2}{1+\overline{R}_1 R_2}\right) \tag{25}$$

This phase resembles the relativistic Thomas precession and the Berry phase and connects the reflection coefficients of the slab when traversed by electromagnetic waves with opposite directions. The graphene-optical analogies defined in the previous section lead to the conclusion that a similar phase factor appears for quantum wavefunctions in graphene if a gated region is traversed by charge carriers in opposite directions.

This similarity is not to be taken for granted because, for a single interface, the optical and quantum reflection coefficients behave differently. Specifically, from (10) it follows easily that the sign of the reflection coefficient of light waves changes if the propagation direction changes, whereas for the quantum wavefunction no such simple relation exists. On the contrary, $r_{21} = -r_{12}^* \exp(i\Phi)$, with $\Phi$ given in (20).

This simple example reminds us that the optical-graphene analogies must be handled with care due to the different natures of the electromagnetic fields and ballistic charge carriers in graphene. In this respect, the optical-graphene analogy is more restrictive than the analogy between ballistic electrons that satisfy the Schrödinger equation and light waves. In the first case we encounter the chiral symmetry of charge carriers, which has no counterpart in classical optics.

**Conclusions**

We have studied in detail the similarities between polarization states of light and ballistic charge carriers in graphene. As a result, the optical equivalent of quantum wavefunctions, Dirac equation, and the effect of an electrostatic potential were found. In a similar way, the quantum analog of the refractive index of light and of the optical composition law of reflection coefficients (and of the relativistic addition of boosts) were derived. However, we also emphasized the difference between the behavior of quantum wavefunctions in graphene and electromagnetic fields, which is evident in the form of the reflection and transmission coefficients at an interface. This difference is related to the chiral symmetry of ballistic charge carriers in graphene, which cannot be mimicked in classical polarization optics. The most striking expression of the inability of classical optics to simulate quantum chirality is the need to impose boundary conditions at different planes in order to recover the quantum reflection coefficient. The results obtained in this paper can, however, provide a basis for the design of optical experiments that mimic the propagation of quantum wavefunctions in graphene. Although the analogies developed here are valid only for ballistic charge carriers, they could be used for developing a more general analogy, valid for disordered graphene. Such quantum-classical analogies proved useful for the study of light propagation in random media (see the review in [3]).

**Figure captions**

Fig. 1  Ballistic wave refraction at an interface between an ungated and a gated region: geometry (top) and potential energy profile (bottom).

Fig. 2  The electromagnetic wave refraction at an abrupt interface (left) should be replaced by the configuration at right to obtain the same reflection coefficient as for ballistic charge carriers in graphene.

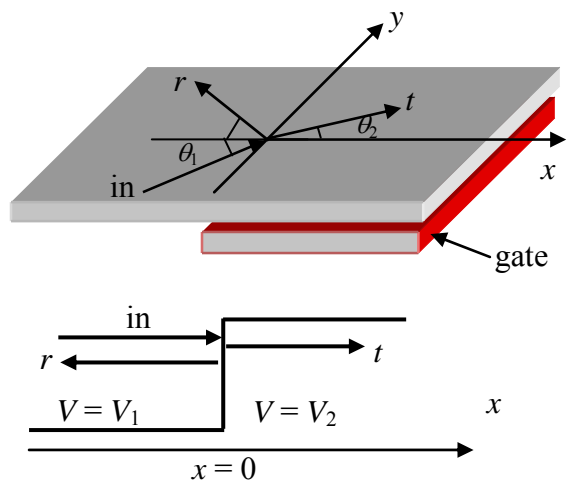

Fig. 1

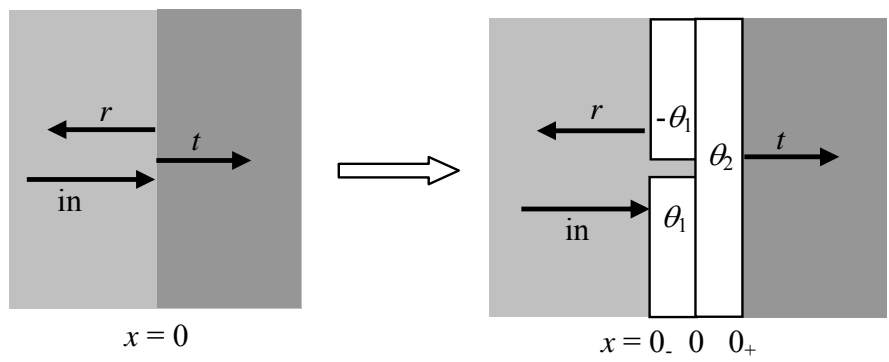

Fig. 2